\documentstyle[12pt]{article}
\title{Stable Quasicrystalline Ground States} 
\author{Jacek Mi\c{e}kisz \\ Institute of Applied Mathematics \\
and Mechanics \\ Warsaw University  \\ ul. Banacha 2  \\ 02-097
Warsaw, Poland \\ e-mail miekisz@mimuw.edu.pl} 
\begin{document} 
\baselineskip=14pt 
\maketitle 
\vspace{30mm}

\noindent {\bf Abstract.} We give a strong evidence that noncrystalline
materials such as quasicrystals or incommensurate solids are not
exceptions but rather are generic in some regions of a phase
space. We show this by constructing classical lattice gas models
with translation-invariant, finite-range interactions and with a
unique quasiperiodic ground state which is stable against small
perturbations of two-body potentials. More generally, we provide
a criterion for stability of nonperiodic ground states.  
\vspace{5mm}

\eject

\section{Introduction} 

One of the important problems in physics, the so-called crystal
problem, is to understand why matter is crystalline at low
temperatures \cite{and,br,sim,uhl1,uhl2,rad1,rad2}. It is
traditionally assumed (but has never been proved) that at zero
temperature, minimization of the free energy of a system of many
interacting particles can only be obtained by their periodic
arrangements (a perfect crystal), which at nonzero temperature
are disrupted by defects due to entropy. Recently, however, there
has been a growing evidence that this basic phenomenon, the
crystalline symmetry of low temperature matter, has exceptions;
in particular incommensurate solids \cite{aub} and, more
recently, quasicrystals \cite{jarr}. It is very important to find
out how generic these examples are. In other words, is 
nonperiodic order present in these systems stable against small 
perturbations of interactions between particles? 
 
The problem of stability of quasiperiodic structures was studied
recently in continuum models of particles interacting through a
well, Lennard-Jones, and other potentials
\cite{wid1,jar1,jar2,jar3,ola,smi}. However, no final conclusion
was reached. After all, one has to compare chosen quasiperiodic
structure with all possible arrangements of particles in the
space, a really formidable task. 
 
Here we will present two classical lattice gas (toy) models with
a unique stable nonperiodic ground state. More precisely, every
site of a square lattice can be occupied by one of several
different particles. The particles interact through two-body
finite-range translation-invariant potentials (we chose chemical
potentials of all particles to be zero). Our models have only
nonperiodic ground-state configurations (infinite-lattice
configurations minimizing the energy density of the system). We
will prove that if one perturbs our models by sufficiently small
chemical potentials or two-body translation-invariant
interactions, their ground-state configurations do not change. It
means that there is an open set in a space of two-body
interactions without periodic ground-state configurations. This
constitutes a first generic counterexample to the crystal
problem. 
 
In Section 2, we discuss general classical lattice gas models
with a unique nonperiodic ground state. We introduce a criterion,
the so-called strict boundary condition, for nonperiodic ground-
state configurations to be stable against small perturbations of
finite-range potentials. In Section 3, we describe main features
of Robinson's nonperiodic tilings of the plane, construct a
classical lattice gas model based on them, and show why its
unique nonperiodic ground state is not stable against small
perturbations of chemical potentials. In Section 4, we present
modifications of Robinson's tilings which allows us to construct 
models with a unique nonperiodic ground state satisfying our
stability criteria. Section 5 contains a short discussion. 
 
\section{Classical Lattice Gas Models and Nonperiodic \- Ground States} 
 
A classical lattice gas model is a system in which every site of
a lattice ${\bf Z}^{d}$ can be occupied by one of $n$ different
particles. An infinite-lattice configuration is an assignment of 
particles to lattice sites, i.e., an element of $\Omega =
\{1,...,n\}^{{\bf Z}^{d}}$. If $X \in \Omega$ and $A \subset {\bf
Z}^{d}$, then we denote by $X(A)$ a restriction of $X$ to $A$.
Particles at lattice sites ${\bf a}$ and ${\bf b}$ interact
through a two-body translation-invariant potential $\Phi ({\bf
a}-{\bf b})$, which is a function on $\{1,...,n\}^{\{{\bf a},{\bf
b}\}}$ - the space of all possible assignments of particles to
lattice sites ${\bf a}$ and ${\bf b}$, and we assume that $\Phi
({\bf a}-{\bf b})=0$ if $dist({\bf a}, {\bf b}) > r$. The
Hamiltonian in a bounded region $\Lambda$ can be then written as 
$H_{\Lambda}=\sum_{{\bf a},{\bf b} \in \Lambda} \Phi ({\bf
a}-{\bf b})$.

$Y$ is a {\bf local excitation} of $X$, $Y \sim X$, $Y,X \in
\Omega$ , if there exists a bounded $\Lambda \subset {\bf Z}^{d}$
such that $X = Y$ outside $\Lambda.$ 

For $Y \sim X$ the {\bf relative Hamiltonian} is defined by 
$$H(Y,X)=\sum_{\{{\bf a},{\bf b}\} \cap \Lambda \neq \emptyset}  
(\Phi ({\bf a}-{\bf b})(Y)-\Phi ({\bf a}-{\bf b})(X)).$$

$X \in \Omega$ is a {\bf ground-state configuration} of $H$ if 
$$H(Y,X) \geq 0 \; \; for \; \; any \; \; Y \sim X.$$  

That is, we cannot lower the energy of a ground-state
configuration by changing it locally.

The energy density $e(X)$ of a configuration $X$ is 
$$e(X)=\liminf_{\Lambda \rightarrow {\bf Z}^{d}}
\frac{H_{\Lambda}(X)}{|\Lambda|},$$
where $|\Lambda|$ is the number of lattice sites in $\Lambda$. It
can be shown that any  ground-state configuration has the minimal
energy density (for a proof see Appendix A). It means that local
conditions present in the definition of a ground-state
configuration force global minimization of the energy density.

Equilibrium behavior of a system of many interacting particles
can be described by a grand-canonical ensemble which tells us
what are probabilities of finding particles at given positions.
An infinite-volume limit of this ensemble is called a
translation-invariant Gibbs state or equilibrium state.
Mathematically speaking, it is a measure on the space, $\Omega$,
of all configurations (for precise definitions see Appendix B).
Then by a {\bf ground state} we mean a limit of a nonzero-
temperature equilibrium state as temperature approaches zero,
with other variables, such as chemical potentials, kept fixed. A
ground state is, therefore, a translation-invariant probability
measure on $\Omega$. It can be shown that the set of all ground-
state configurations have probability one with respect to a
ground-state measure.

If a system has a unique periodic ground-state configuration and
its translations, then a unique ground-state measure assigns an
equal probability to all these translations. For example, the
Ising antiferromagnet has two alternating ground-state
configurations but only one ground-state measure which assigns
probability $1/2$ to both of them.   

Generally, a probability ground-state measure $\mu$ gives equal
weights to all ground-state configurations and can be obtained as
a limit of averaging over a given ground-state configuration $X$
and its translations $\tau_{{\bf a}}X$ by  lattice vectors ${\bf
a} \in {\bf Z}^{d}$: $\mu= \lim_{\Lambda \rightarrow {\bf Z}^{d}}
\frac{1}{|\Lambda|} \sum _{{\bf a} \in \Lambda} \delta
(\tau_{{\bf a}} X)$, where $\delta (\tau_{{\bf a}}X)$ is a
probability measure assigning probability $1$ to $\tau_{{\bf
a}}X$.

For any potential, the set of ground-state configurations is
nonempty but it may not contain any periodic configurations
\cite{rad3,rad4,rad5,mier,mie1,mie2,mie3,rad6}. We restrict
ourselves to systems in which, although all ground-state 
configurations are nonperiodic, there is a unique translation-invariant
probability measure supported by them. Therefore, such measure is
inevitably a zero-temperature limit of translation-invariant
Gibbs states (equilibrium states) and hence it is a ground state 
of a given model.
Uniqueness of a ground-state measure is equivalent to the
statement that a uniformly defined frequency of any finite
arrangement of particles is the same in all ground-state
configurations (such measures are called uniquely ergodic
measures and their supports are called uniquely ergodic sets) \cite{oxt}.
As a support of a ground-state measure we will always choose
a minimal set (if any arrangement appears in a ground-state
configuration, it appears with a positive density), i.e., we 
exclude ground-state configurations with broken bonds (such as
interface ground-state configurations of the ferromagnetic Ising model).
More precisely, to find the frequency
of a finite arrangement in a given configuration we first count
the number of times it appears in a box of size $l$ which is
centered at the origin of the lattice, divide it by $l^{d}$, and
then take the limit $l \rightarrow \infty$. If the convergence is
uniform with respect to the position of the boxes, then we say
that the configuration has a uniformly defined frequency of this
arrangement.  

Let us emphasize again that in our models the frequency of any
finite particle arrangement is the same in all ground-state
configurations. Yet, it is not true that all ground-state
configurations are lattice translations of a single nonperiodic
ground-state configuration. That is why, in order to deal with
all ground-state configurations simultaneously, we have to
introduce a ground-state measure. Such a situation is present in
all tiling models of quasicrystals; for example in the Penrose
tilings, where instead of ground-state configurations we have
perfect tilings. 

To get all ground-state configurations we have to close (in a
topology described in Appendix B) a set of all translations of
any ground-state configuration (the closure does not depend on
the choice of a configuration). In this way we obtain uncountably
many ground-state configurations. In fact, probability of any
single configuration should be zero; otherwise, the measure of a
countable union of its translations would be infinite. If there
were countably many ground-state configurations (there are
countably many translations), then the measure of the set of all
ground-state configurations would be zero, and this is
impossible.

We will introduce now a condition which is equivalent to the
stability of nonperiodic ground states. It generalizes the so-
called Peierls condition \cite{sinb} for models without periodic
ground-state configurations.

For clarity of presentation we assume that our models are two-
dimensional. We also assume that our models have a unique ground state,
supported by ground-state configurations for which
all interactions attain simultaneously their
minima (we choose them to be equal to zero). As it was explained before,
we do not have to consider ground-state configurations with broken bonds.
Therefore, if $Y$ is
not a ground-state configuration, it contains pairs of particles
with nonminimal energies (we choose them to be equal to $1$), the
so-called {\bf broken bonds}. Denote by $B(Y)$ the number of
broken bonds in $Y$. Thus $H(Y,X)=B(Y)$ if $Y \sim X$ and $X$
is a ground-state configuration. 
\newtheorem{prop}{Proposition}
\newtheorem{theo}{Theorem}
\vspace{2mm} 

\noindent {\bf Condition} The strict boundary condition for local excitations.
\vspace{2mm}

\noindent Let $X$ be a ground-state configuration and $Y$ a local
excitation of $X$; $Y \sim X$. Let $n_{ar}(Y,X)$ denote the
difference of the number of appearances of an arrangement $ar$ (a
particle or a pair of particles) in $Y$ and the number of
its appearances in $X$.

We say that a model satisfies the strict boundary condition for
local excitations and an arrangement $ar$ if
there exists a $C_{ar}>0$ such that for every ground-state
configuration $X$ and every $Y \sim X$
\begin{equation}
|n_{ar}(Y,X)| <C_{ar}B(Y).
\end{equation}

\begin{theo}
A unique ground state of a finite-range Hamiltonian, $H$, is
stable against small perturbations of chemical potentials and
two-body interactions of range smaller than $r$ if and only if
the strict boundary condition is satisfied for particles and
pairs of particles at a distant smaller than $r$.
\end{theo}
{\bf Proof:} Assume first that the strict boundary condition
does not hold for a
ground-state configuration $X$ and an arrangement $ar$ (a
particle or a pair of particles at a distance smaller than $r$).
Hence, for any $C>0$ there exists $Y_{C} \sim X$ such that
$n_{ar}(Y_{C},X)>CB(Y_{C})$ or $n_{ar}(Y_{C},X) < - CB(Y_{C})$.
We assume that the first case holds; the second one can
be treated in an analogous way. We introduce a 
perturbation, $\Psi$, which assigns a negative energy, $E=-1/C$,
to the arrangement $ar$. This perturbation is small if $C$ is large.
Let $H''=H+ H'$, where
$H'=\sum_{\Lambda} \Psi_{\Lambda}$. Then
\begin{equation}
H''(Y_{C},X)=B(Y_{C})-n_{ar}(Y_{C},X)/C < 0.
\end{equation}
Thus, $X$ is not a ground-state configuration for the perturbed
Hamiltonian $H''$. 

Assume now that $|n_{ar}(Y,X)|<C_{ar}B(Y)$ for every arrangement
$ar$ (a particle or a pair of particles at a distance smaller
than $r$) and every local excitation $Y \sim X$ of a ground-state
configuration $X$. Let $\Psi$ be a potential of a range smaller
than $r$ and $H''= H + H'$ be a perturbed Hamiltonian, where $H'
= \sum_{\Lambda}\Psi_{\Lambda}$. Then
\begin{equation}
H''(Y,X)=B(Y) + \sum_{ar} n_{ar}(Y,X) \Psi (ar)
\end{equation}
\begin{equation}
> B(Y)-B(Y)\sum_{ar}C_{ar}|\Psi (ar)| >0
\end{equation}
if
\begin{equation}
\sum_{ar}C_{ar}|\Psi (ar)|<1.
\end{equation} 
Hence, $X$ is a ground-state configuration for $H''$ for every sufficiently
small perturbation.

So far we have proved that ground-state
configurations of $H$ (without broken bonds) are ground-state configurations
of $H''$. Now we will show that the set of ground-state configurations of 
$H''$ is a uniquely ergodic set. It will then follow that the ground state
of $H''$ must be equal to that of $H$. 
Let $\mu''$ be an ergodic measure different from the ground-state
measure of $H$. It has therefore a nonzero density, $\omega$,
of broken bonds.
Let $Z$ be in the support of
$\mu''$ and have the density $\omega$ of
broken bonds. We will now show that $Z$ is not a ground-state
configuration of $H''$.

We construct two configurations: $W \sim Z$ such that $W=X$  on a
square $A$ of size $l$ and $W=Z$ outside $A$, and $Y \sim X$ such
that $Y=Z$ on $A$ and $Y=X$ outside $A$.

Since $Y$ is an excitation of $X$, then
\begin{equation}
|n_{ar}(Y,X)|<C_{ar}B(Y)<C_{ar}(\omega l^{2} + o(l^{2})).
\end{equation}
When we change $Z$ to $X$ on $A$ we decrease the number of broken
bonds at least by $\omega l^{2}-o(l^{2})$ and change the number
of appearances of $ar$ at most by $C_{ar}(\omega l^{2}+o(l^{2}))$ which
follows from (7). Hence,
\begin{equation}
H''(W,Z) < -(\omega l^{2} - o(l^{2})) + \sum_{ar}|\Psi
(ar)|(C_{ar}(\omega l^{2} + o(l^{2})).
\end{equation}
Thus,
\begin{equation}
H''(W,Z)<0
\end{equation}
if (6) holds and $l$ is big enough. It follows that $Z$ is not a
ground-state configuration of $H''$. $\Box$

\section{Robinson's Tilings and Unstable Ground States} 
 
To construct our classical lattice gas models, we will use
Robinson's nonperiodic tilings of the plane \cite{rob,gr}. He
designed 56 square-like tiles such that using an infinite number
of their copies one can tile the plane only in a nonperiodic
fashion. We describe now the Robinson tiles; we follow \cite{rob}
closely. There are five basic tiles represented symbolically in
Fig.1; all other tiles can be obtained from them by rotations and
reflections. The first tile on the left is called a cross; the
other tiles are called arms. Every tile is also furnished with
one of the four parity markings shown in Fig.2. Crosses can be
combined with the parity marking at the lower left in Fig.2.
Vertical arms (the direction of long arrows) can be combined with
the marking at the lower right and horizontal arms with the
marking at the upper left. All tiles can be combined with the
remaining marking. Two tiles ``match'' if arrow heads meet arrow
tails separately for the parity markings and the markings of
crosses and arms. 
 
We will now describe main features of Robinson's nonperiodic
tilings. Observe first that, in any tiling, the centers of the
tiles form the square lattice ${\bf Z}^{2}$. Therefore, any
tiling can be described by an assignment of tiles to lattice
sites, a lattice configuration. It is also easy to see that if
the plane is tiled with tiles with the above parity markings,
then these must alternate both horizontally and vertically in the
manner shown in Fig.2. Hence, every odd-odd position on the ${\bf
Z}^{2}$ lattice (if columns and rows are suitably numbered) is
occupied by crosses with the lower left parity marking. They form
a periodic configuration with period $4$ as shown in Fig.3
(crosses are denoted there by $\lfloor$, where line segments
denote double arrows). Then in the center of each ``square''
matching rules force a cross such that the previous pattern
reproduces but this time with period $8$.

Continuing this procedure (the so-called principle of expanding
squares) infinitely many times, we obtain a nonperiodic
configuration. It has built in periodic configurations of period
$2^{n}$, $n\geq 2$, with tiles on sublattices of ${\bf Z}^{2}$ as
is shown in Fig.4. Each side of every square consists of two
lines of arms with long double arrows at the left of Fig.1
(called double arms) meeting half-way through double short arrows
of arms at the top of Fig.1. It is exactly the arms with
asymmetrically located double short arrows which force crosses to
be present at the center of each square. Crosses of different
squares at the same sublattice are joined by lines of arms with
single long arrows at the right of Fig.1 (called single arms).
For a complete construction of Robinson's tilings and proofs, we
refer the reader to the original article of Robinson \cite{rob}. 

Now, using Robinson's tiles we will construct a lattice gas model
\cite{rad3,rad4,rad5,mier,mie1,mie2,mie3,rad6}. Every site on a
square lattice can be occupied by one of the particles
corresponding to the tiles. Two nearest-neighbor particles which
do not match as tiles have a positive interaction energy, say
$1$; otherwise, the energy is zero. We obtain a lattice gas model
with nearest-neighbor translation-invariant interactions.

Any periodic configuration with period $p$ corresponds to a
periodic configuration of tiles. Therefore, it has, in any square
of $p^{2}$ lattice sites, at least one pair of nearest neighbors
with energy equal to $1$. Hence, any periodic configuration has a
nonzero energy density. On the other hand, nonperiodic
configurations corresponding to tilings have energy density equal
to zero. It follows that the model does not have periodic ground-
state configurations.

It has been proven \cite{mier} that the unique ground state of
this model is not stable against small perturbations of chemical
potentials. To see this let us introduce a small negative
chemical potential $h$ for single arms. Notice that every ground-
state configuration contains arbitrarily long sequences of double
arms. When we change such a sequence of length $l$ into a
sequence of single arms, we lower the energy by $lh$ along the
sequence, and only at two endpoints of the sequence the energy
increases by $1$. Hence, no matter how small $h$ is, one can
always change locally a ground-state configuration of an
unperturbed model to lower its energy. It is easy to see that
the strict boundary condition is not satisfied.

In fact, we showed that frequencies of single and double arms
change continuously with respect to their chemical potentials.
This behavior of continous change in stoichiometries at zero
temperature is not present in previously known models and real
alloys. 
 
\section{Stable Quasiperiodic Ground States} 
 
We will present in a moment models with a unique nonperiodic
ground state satisfying our stability criteria. This will be a
modification of the model based on Robinson's tiles such that a
sufficiently small chemical potential for one kind of arm and a
small two-body perturbation do not change its unique ground
state.

We cannot simply get rid of the double arms because it is exactly
the double arms which create squares whose central crosses are
forced. The absence of double arms will destroy the principle of
expanding squares and we will no longer have a unique nonperiodic
ground state.

However, let us recall that the forcing is done by four arms
located exactly at the middle point between crosses forming a
square; these are the arms at the top of Fig.1. The main idea of
our construction is to force similar arms in the middle of every
side of every square, as in the original Robinson tiling, by
using periodic sequences of new markings instead of translation-
invariant sequences of single and double arms. We will use
sequences of alternating $A$, $B$, and $C$ markings propagating
from crosses. Whenever two $B$ or $C$ markings meet at the middle
of a side of a square, then our special arms are forced there. 

Observe that we do not have long sequences of different arms
anymore, just three translations of lines $...ABCABC...$. It is
impossible to distinguish between these translations by chemical
potentials and also by two-body interactions because of the
symmetry of the tiles and the fact that $2^{n}$ ``periods'' of
Robinson's tilings and period $3$ of the sequences are relatively
prime.
\vspace{3mm}
 
\noindent {\bf MODEL I} 
\vspace{3mm}
 
\noindent We will describe first our tiles. These are modified crosses and
arms with the parity markings as before. We have also two
additional levels of letter markings. Each side of each tile is
marked by $A$, $B$, or $C$ and by $T$, $O$, $M$, $E$, or $K$ (not
all choices are present as explained below). All possible
markings of crosses are shown in Fig.5. Next we allow all
rotations of crosses but we do not allow any reflections. This
feature makes our model somewhat different from the original
Robinson one. We have $15$ types of arrows present in arms: all
combinations of $A \rightarrow B$, $B \rightarrow C$, $C
\rightarrow A$ and $T \rightarrow O$, $O \rightarrow M$, $M
\rightarrow E$, $E \rightarrow K$, $K \rightarrow T$. Line
segments present in arms are marked $A-A$, $B-B$, or $C-C$ while
$B-B$ and $C-C$ segments are also marked $T-O$ or $O-T$ as is
shown in Fig.6. Again, we allow rotations but not reflections.

Two tiles match, if in addition to previous requirements, letter
markings match separately on both levels. We also assume that
line segments $A-A$ can be put next to any letter marking of the
second level (this is not a tiling condition).

Our matching rules can be translated as before into a lattice gas
model with nearest-neighbor interactions. In addition, we have to
take care of the fact that letter markings of the second level
which are opposite nearest neighbors of $A-A$ line segments are
not forced to be correlated. We introduce a two-body interaction
of range $2$ between two arrows facing one another such that the
energy is zero if one of them has $T$ marking and 
the other one is marked by $O$; otherwise, the energy is $1$. 
 
We will show that the unique ground-state measure of the above
model is of the Robinson type. More precisely, looking at crosses
and ignoring $A$ arrows on every sublattice $2^{n}{\bf Z}^{2}$
with $n$ odd and $C$ arrows for $n$ even, we see the self-similar
structure of expanding squares shown in Fig.4. 
  
\begin{prop} 
The unique ground-state measure of Model I is of the Robinson
type as described above. 
\end{prop} 
{\bf Proof:} First, let us recall that the parity markings force
crosses to occupy every site of a $2{\bf Z}^{2}$ sublattice.
Letter markings of the second level allow only crosses which are
rotations of the cross at the upper left in Fig.5.

Arms with the upper left and lower right parity marking force
relative orientations of these crosses to be such that squares of
length $2$ are created as is shown in Fig.3 (arrows of crosses
with $A$ markings and arms are not drawn and remaining arrows are
represented by line segments). The above arms with $C-C$ line
segments force crosses to appear in the center of every square 
just created. New crosses (four rotations of the cross at the
upper right in Fig.5) form a $4{\bf Z}^{2}$ sublattice.

Once we have crosses on a sublattice, lines of arrows propagate
from every cross and meet half-way through line segments of arms.
For the crosses on a $2^{n}{\bf Z}^{2}$ sublattice with $n$ odd,
these are alternating $C-C$ and $A-A$ line segments and
alternating $B-B$ and $A-A$ segments for $n$ even. These segments
force crosses to have relative orientations on every sublattice
as is shown in Fig.3. $C-C (B-B)$ segments force crosses in
centers of squares just created. Crosses on successive
sublattices have to be chosen in clockwise order from Fig.5. The
upper left corner of every square has to be occupied by an
appropriate cross from Fig.5 and other corners by clockwise
rotations. An infinite nonperiodic configuration is inductively
created. $\Box$ 
  
\begin{prop} 
Model I satisfies the strict boundary condition for
particles and pairs of particles, if we do not distinguish
between their different rotations and reflections. 
\end{prop} 
{\bf Proof:} Let a broken bond be a unit segment on the dual
lattice separating two nearest-neighbor particles with a positive
interaction energy (a common side of two nearest-neighbor tiles
which do not match) or separating two arrows at a distance $2$
with the positive energy. Let us then divide an excitation $Y$ of
a ground-state configuration $X$ into connected components
without broken bonds (two lattice sites are connected if they are
nearest neighbors) such that on every component crosses having 
relative orientations shown in Fig.3 form a $2{\bf Z}^{2}$
sublattice. This is achieved by paths on the lattice dual to
${\bf Z}^{2}$ with lengths not bigger than $16$ and joining
broken bonds.

When one considers only $2{\bf Z}^{2}$ sublattices and lines of
arms joining their sites, the absolute value of the difference
between the number of appearances of a given particle in $Y$ and
the number of its appearances in $X$ is not bigger than $16$ times
the number of the above paths.

Now we decompose  every connected component into components with
crosses that form $4{\bf Z}^{2}$ sublattices. This is achieved by
paths on the dual lattice with lengths not bigger than $32$ and
again joining broken bonds. When one considers $4{\bf Z}^{2}$
sublattices and lines of arms joining their sites, the additional
difference of the number of appearances of our particle is not
bigger than $16$ times the number of new paths plus $16/2$ times
the number of old paths.

At the $n-th$ step we divide every connected component
constructed so far into components with crosses forming
$2^{n}{\bf Z}^{2}$ sublattices and we use paths with lengths not
bigger than $2^{n+3}$. Again, the additional difference of the
number of appearances of our particle is bounded by $16$ times
the number of new paths plus $1/2$ of the difference at the
$(n-1)th$ step.

We repeat this process $m$ times, where $m$ is the smallest
number such that $diam \{{\bf a}, Y({\bf a}) \-\neq X({\bf a})
\}<2^{m}$. Now, broken bonds and paths can be regarded as
vertices and edges of a planar graph. It follows from Euler's
formula that the total number of all paths is bounded by $3$
times the number of broken bonds. This shows that $|n(Y,X)| \leq
96B(Y)$ for a given particle.

We will now discuss two-particle arrangements. For a pair of
particles at a distance $D$, at least one particle is located on
a sublattice $2^{k}{\bf Z}^{2}$ with $2^{k} \leq 2D$ or on a line
joining sites of this sublattice. To get a bound for a pair of
particles we must take into account the effect of shifting one
line of arms joining two sites of a sublattice with respect to
the other parallel line. This can be done by multiplying the
previous bound by $2D$.      $\Box$ 
 
Let us now perturb our model. We will consider only covariant
interactions. That is, a chemical potential of a particle-tile or
a two-body interaction between a pair of particles should be the
same for any of their rotations and reflections. Combining
Theorem 1 and Proposition 2 we get 
 
\begin{theo} 
The unique nonperiodic ground state of Model I is the unique
ground state for any sufficiently small chemical potential and
two-body covariant perturbation. 
\end{theo} 
 
\noindent {\bf MODEL II}
\vspace{3mm} 
 
\noindent Let us now describe tiles of our second model. We have modified
crosses with $A$ and $C$ arrows shown in Fig.7, arms with $A
\rightarrow B$, $B \rightarrow C$, and $C \rightarrow A$ arrows
and $A-A$, $B-B$, and $C-C$ line segments shown in Fig.8 and 9.
Arms in Fig.8 are combined with the upper right parity marking.
Arms in Fig.9 can be combined with the lower right or upper left
parity marking. Crosses are combined with the parity markings as
before. We allow all rotations and reflections of our new tiles
as in the original Robinson model. 
 
We do not have anymore the previous second level of letter
markings. However, $B-B$ and $C-C$ line segments are not located
in the middle of arms anymore so they force crosses to create
squares. Arms with central $A-A$ segments from Fig.8 cannot
prevent these squares from going out of phase; a fault shown in
Fig.10 can be created. Such faults are also present in some of
the original Robinson tilings. One may prove that any tiling has
at most one fault (follow the proof of the analogous theorem in
\cite{rob}, look also at the proof of Proposition 3). Therefore,
as in the previous model, the unique ground state of Model II is
of the Robinson type. 
 
\begin{prop} 
Model II satisfies the strict boundary condition for particles and pairs of
nearest-neighbor particles, if we do not distinguish between 
their different rotations and reflections. 
\end{prop} 
{\bf Proof:} Observe first that arms with the lower right and the
upper left parity marking have asymmetric $A-A$ line segments.
Therefore, there are no faults on $2{\bf Z}^{2}$ sublattices.
The first step is the same as in the proof of Proposition 2.

In the second step we decompose every connected component without
broken bonds into components with crosses that form $4{\bf
Z}^{2}$ sublattices but not necessarily having relative
orientations as in Fig.3; faults may be present. This is
achieved by paths on the dual lattice with lengths not bigger
than $32$ and joining broken bonds.

We may have an arbitrarily long faults perpendicular to $A-A$
line segments without any broken bonds except at the endpoints of
the fault. The presence of two arms on the fault with line
segments parallel to the fault forces a cross (or a broken bond)
between them and then a broken bond is forced at a distance not
bigger than $16$ from the cross. Arms with $B-B$ or $C-C$ line
segments perpendicular to the fault also force a cross on the
fault or a broken bond at a distance not bigger than $16$. Now,
when one considers only $4{\bf Z}^{2}$ sublattices and lines of
arms joining their sites, then the absolute value of the
difference between the number of appearances of a given particle in an
excitation $Y$ and the number of its appearances in a ground-state
configuration $X$ is not bigger than $16$ times the number of
paths in the present decomposition plus $16/2$ times the number
of paths in the first step (there is no change of the number of
particles and pairs of nearest neighbors along a fault with
$A-A$ line segments perpendicular to it).

Now we decompose every connected component into components with
crosses that form $8{\bf Z}^{2}$ sublattices. These crosses go
out of phase along previous faults and they may also create they
own faults. We repeat this process $m$ times, as in the proof of
Proposition 2, to get $|n(Y,X)|<96B(Y)$ for particles and pairs
of nearest-neighbor particles.  $\Box$ 
 
\begin{theo} 
The unique nonperiodic ground state of Model II is the unique
ground state for any sufficiently small chemical potential and
nearest-neighbor covariant perturbation. 
\end{theo} 
 
Let us notice that particles of our models have shapes and
interactions between them are covariant but not symmetric: an
interaction between nearest neighbors $RP$ is the same as between
\hspace{8mm} but not necessarily the same as between $PR$. There
is, however, a construction due to Radin \cite{rad4} who enlarges
a family of possible particles and makes them shapeless, and
interactions between them fully symmetric (dependent only on the
distance between given particles). He constructs in this way a
classical lattice gas model with a finite-range fully symmetric
interactions and with a unique nonperiodic ground state. His
method requires the original family of particles-tiles to be both
rotation and reflection-invariant. Therefore, it can be applied 
only to our second model. We use it to construct a model with
shapeless particles and with a unique nonperiodic ground state 
which is stable in a space of fully symmetric interactions. 
 
\begin{theo} 
The unique ground state of the symmetric version of Model II is
the unique ground state for any sufficiently small chemical
potential and symmetric perturbation of range smaller than three
lattice spacings. 
\end{theo} 
 
\section{Conclusions} 
 
We have constructed two finite-range classical lattice gas models
with a unique quasiperiodic ground state which is stable against
small perturbations of interactions.

The family of particles of our first model is rotation-invariant
but not reflection-invariant. Our quasiperiodic ground state is
stable in the space of covariant two-body interactions.

In the second model, particles are shapeless and interaction
between them depends only on the distance between them. The
unique quasiperiodic ground state is stable in the space of fully
symmetric interactions of range smaller than three lattice
spacings. 

In the construction of our models we used the presence of order
in nonperiodic ground-state configurations. Although nonperiodic,
they are not, however, completely chaotic. They exhibit
long-range positional order in the sense that arrangements of
particles at distant regions are correlated. In fact, our ground-
state configurations possess highly ordered structures: if a
certain fraction of particles is ignored, the rest of a ground
state configuration is periodic; the smaller the fraction, the
larger the period. It would be interesting to see if other, more
disordered, ground states are stable. Modifying the principle of
expanding squares responsible for nonperiodicity of Robinson's 
tilings, many models without periodic ground states were recently
constructed \cite{moz,rad6,tsi}.

Let us point out that the Robinson tilings share with the Penrose tilings
many essential features like forced nonperiodicity, self-
similarity, uniqueness of a structure for given matching rules.
In fact, using the so-called Ammann bars, the Penrose tilings can
be also represented by configurations on a square lattice
\cite{gr}. We hope that investigating Robinson's tilings one may
provide some information about behavior of quasicrystals
associated with the Penrose tilings. Let us also note that models
based on the Penrose tilings and with particles on a square
lattice were recently investigated in Ref. \cite{wid2}.
  
Finally, let us say few words about positive temperatures, where
to construct stable equilibrium phases of a system of many
interacting particles one has to minimize not the energy but the
free energy of the  system taking into account disruptions due to
entropy. For lattice gas models with a finite number of periodic
ground-state configurations satisfying the so-called Peierls
condition, there exists a complete theory due to Pirogov and Y.Sinai
\cite{sin1,sin2,sinb}. It tells us that for every Hamiltonian
with the above properties, a low-temperature phase diagram is a small
deformation of a zero-temperature phase diagram. We would like to generalize
the Pirogov-Sinai theory to be able to apply it to lattice models
without any periodic ground-state configurations. One of the important
problems is to construct a
quasiperiodic equilibrium phase for a finite-range interaction -
a microscopic model of a quasicrystal. We conjecture the
existence of quasiperiodic equilibrium phases in a modified
three-dimensional version of our second model. 

\section{Appendix A}
\begin{theo}
Let $X$ be a ground-state configuration of a finite-range
Hamiltonian $H$. Then $X$ has the minimal energy density: $e(X)
\leq e(Y)$ for any $Y \in \Omega$.
\end{theo}
{\bf Proof by contraposition:} For clarity of presentation we
consider two-dimensional models. Assume that there exists a
configuration $Y$ with smaller energy density than that for a
ground-state configuration $X$: $e(Y) < e(X)$. Then there is a
square, $Sq(N)$, of $N^{2}$ lattice sites and $\delta >0$ such
that
\begin{equation}
H_{Sq(N)}^{\Phi}(Y) < H_{Sq(N)}^{\Phi}(X) - N^{2}\delta
\end{equation}
and 
\begin{equation}
8N\sum_{0 \in 
\Lambda}max_{W \in \Omega_{Lambda}}|\Phi_{\Lambda}(W)|<N^{2}\delta.
\end{equation}
Now we construct an excitation $Z \sim X$ as follows:
$Z(Sq(N))=Y(Sq(N))$ and $Z=X$ outside $Sq(N)$. It follows that
$H(Z,X)<0$ so $X$ is not a ground-state configuration.  

\section{Appendix B}
We equip $\{1,...,n\}$ with the discrete topology (every subset
of $\{1,...,n\}$ is both open and closed). $\Omega$ is then
compact in the product topology. Let $ar$ be a finite arrangement
of particles, $ar \in \{1,...,n\}^{\Lambda}$ for some bounded
$\Lambda$, then $$C^{ar}_{\Lambda}=\{X \in \Omega:
X(\Lambda)=ar\}$$
is a {\bf cylinder set} based in $\Lambda$. Cylinder sets are
open and they generate all open sets in $\Omega$ (every open set
is a union of some cylinders). It follows that a sequence of
configurations $X_{m}$ converges to $X$, 
$$X_{m} \rightarrow X \; as \; m \rightarrow \infty ,$$
if for every bounded $\Lambda \in {\bf Z}^{d}$ there is
$m(\Lambda)$ such that for every $m > m(\Lambda)$,
$X_{m}(\Lambda)=X(\Lambda)$.  

{\bf Borel} sets are generated from open sets by taking
complements and countable unions. 

A {\bf probability measure}, $\mu$, on a configuration space,
$\Omega$, is a function which assigns a number between $0$ and
$1$ to every Borel set such that $\mu(\Omega)=1,$
$\mu(\emptyset)=0,$ and
$$\mu(\cup_{i=1}^{\infty}A_{i})=\sum_{i=1}^{\infty}\mu(A_{i})$$
for Borel sets $A_{i}$ and $A_{j}$ if $A_{i}\cap A_{j} =
\emptyset$ for all $i \neq j.$ 
\vspace{5mm}

\noindent {\bf Acknowledgments.} I have begun this research at Institut de
Physique Th\'{e}orique, Unit\'{e} FYMA, Catholique Universit\'{e}
de Louvain-la-Neuve (it was supported there by Bourse de
recherche UCL/FDS) and continued at Instituut voor Theoretische
Fysica, Katholieke Universiteit Leuven. I would like to thank the Polish 
Committee for Scientific Research - KBN for a financial support under
the grant 2P03A01511.


\begin{thebibliography}{99} 
\bibitem{and} P. W. Anderson, {\em Basic Notions of Condensed
Matter Physics} (Benjamin, Menlo Park, 1984), p.12. 
\bibitem{br} S. G. Brush, {\em Statistical Physics and the Atomic
Theory of Matter, from Boyle and Newton to Landau and Onsager}
(Princeton University Press, Princeton, 1983), p.277. 
\bibitem{sim} B. Simon, in {\em Perspectives in Mathematics:
Anniversary of Oberwolfach 1984} 
(Birkhauser-Verlag, Bassel, 1984), p.442. 
\bibitem{uhl1} G. E. Uhlenbeck, in {\em Fundamental 
Problems in Statistical Mechanics II}, edited by E. G. D. Cohen
(Wiley, New York, 1968), p.16.
\bibitem{uhl2} G. E. Uhlenbeck, in {\em Statistical Mechanics;
Foundations and Applications}, edited by T. A. Bak (Benjamin, New
York, 1967), p.581. 
\bibitem{rad1} C. Radin, {\em Low temperature and the origin of
crystalline symmetry,} {\em Int. J. Mod. Phys.} {\bf B1}: 1157
(1987). 
\bibitem{rad2} C. Radin, {\em Global order from local sources,}
{\em Bull. Amer. Math. Soc.} {\bf 25}: 335 (1991). 
\bibitem{aub} S. Aubry, {\em Devil's staircase and order without
periodicity in classical condensed matter,} {\em J. Physique}
{\bf 44}: 147 (1983). 
\bibitem{jarr} {\em Aperiodicity and Order} Vol 1 and 
2, edited by M. V. Jari\'{c}  (Academic Press, London, 1987,
1989). 
\bibitem{wid1} M. Widom, K. J. Strandburg, and R. H. Swendsen,
{\em Quasicrystal ground state,} {\em Phys. Rev. Lett.} {\bf 58}:
706 (1987). 
\bibitem{jar1} S. Narasimhan and M. V. Jari\'{c}, {\em
Icosahedral quasiperiodic ground states?} {\em 
Phys. Rev. Lett.} {\bf 62}: 454 (1989). 
\bibitem{jar2} A. P. Smith and D. A. Rabson, {\em  Comment on
"Icosahedral quasiperiodic ground states?"} {\em 
Phys. Rev. Lett.} {\bf 63}: 2768 (1989). 
\bibitem{jar3} S. Narasimhan and M. V. Jari\'{c} reply {\em Phys.
Rev. Lett.} {\bf 63}: 2769 (1989). 
\bibitem{ola} Z. Olami,  {\em Stable dense icosahedral
quasicrystals,} {\em Phys. Rev. Lett.} {\bf 65}: 2559 (1990). 
\bibitem{smi} A. P. Smith, {\em Stable one-component
quasicrystals,} {\em Phys. Rev. B} {\bf 43}: 11635 (1991). 
\bibitem{rad3} C. Radin, {\em Tiling, periodicity, and crystals,}
{\em J. Math. Phys.} {\bf 26}: 1342 (1985). 
\bibitem{rad4} C. Radin, {\em Crystals and quasicrystals: a
lattice gas model,} {\em Phys. Lett.} {\bf 114A}: 381 (1986). 
\bibitem{rad5} C. Radin, {\em Crystals and quasicrystals: a
continuum model,} {\em Commun. Math. Phys.} {\bf 105}: 385
(1986). 
\bibitem{mier} J. Mi\c{e}kisz and C. Radin, {\em The unstable
chemical structure of the quasicrystalline alloys,} {\em Phys.
Lett.} {\bf 119A}: 133 (1986). 
\bibitem{mie1} J. Mi\c{e}kisz, {\em Many phases in systems
without periodic ground states,} {\em Commun. Math. Phys.} {\bf
107}: 577 (1986). 
\bibitem{mie2} J. Mi\c{e}kisz, {\em Classical lattice gas model
with a unique nondegenerate but unstable periodic ground state
configuration,} {\em Commun. Math. Phys.} {\bf 111}: 533 (1987). 
\bibitem{mie3} J. Mi\c{e}kisz, {\em A microscopic model with
quasicrystalline properties,} {\em J. Stat. Phys.} {\bf 58}: 1137
(1990). 
\bibitem{rad6} C. Radin,  {\em Disordered ground states of
classical lattice models,} {\em Rev. Math. Phys.} {\bf 3}: 125
(1991). 
\bibitem{oxt} J. C. Oxtoby, {\em Ergodic sets,}
{\em Bull. Amer. Math. Soc.} {\bf 58}: 116 (1952).
\bibitem{sin1} Pirogov and Y. G. Sinai, {\em Phase diagrams of classical
lattice systems I}, {\em Theor. Math. Phys.} {\bf 25}: 1185 (1975).
\bibitem{sin2} Pirogov and Y. G. Sinai, {\em Phase diagrams of classical
lattice systems II}, {\em Theor. Math. Phys.} {\bf 26}: 39 (1976).
\bibitem{sinb} Ya. G. Sinai, {\em Theory of Phase Transitions:
Rigorous Results} (Pergamon Press, Oxford, 1982).
\bibitem{rob} R. M. Robinson, {\em Undecidability and
nonperiodicity for tilings of the plane,} {\em Invent. Math.}
{\bf 12}: 177 (1971). 
\bibitem{gr} B. Gr\"{u}nbaum and G. C. Shephard, {\em Tilings and
Patterns} (Freeman, New York, 1986). 
\bibitem{moz} S. Mozes, {\em Tilings, substitutions, and
dynamical systems generated by them,} {\em J. d'Analyse Math.}
{\bf 53}: 139 (1989). 
\bibitem{tsi} B. Tsirelson, {\em A robust nonperiodic tiling
system,} preprint of Tel Aviv 
University, School of Mathematical Sciences (1992). 
\bibitem{wid2} W. Li, H. Park, and M. Widom, {\em Phase diagram
of a random tiling quasicrystal,} {\em J. Stat. Phys.}  {\bf 66}:
1 (1992). 
\end{thebibliography}
\end{document}